\documentclass[apj]{emulateapj}
\usepackage{apjfonts}
\usepackage{ulem} 
\usepackage{color} 
\newcommand{\delR}{\Delta R}
\newcommand{\delK}{\delta k}

\newcommand{\Emax}{E_\mathrm{max}}

\newcommand{\muG}{$\mu$G}
\newcommand{\kpeak}{k_0}
\newcommand{\ETeV}{E_\mathrm{TeV}}

\renewcommand{\vec}[1]{\mathbf{#1}}
\newcommand{\dTobs}{\Delta t_\mathrm{obs}}

\newcommand{\SCly}{self-consistently}
\newcommand{\MC}{Monte Carlo}
\newcommand{\rel}{relativistic}
\newcommand{\syn}{synchrotron}
\newcommand{\NL}{nonlinear}

\newcommand{\rmg}{R_{\rm max}}
\newcount\listnorom
\newcommand\listromanDE{\global\advance \listnorom by 1
{\lowercase\expandafter{(\romannumeral\listnorom)}\ }}
\newcommand\newlistroman{\listnorom=0}

\def\be{\begin{eqnarray}}
\def\ee{\end{eqnarray}}
\def\lsim{\;\raise0.3ex\hbox{$<$\kern-0.75em\raise-1.1ex\hbox{$\sim$}}\;}
\def\gsim{\;\raise0.3ex\hbox{$>$\kern-0.75em\raise-1.1ex\hbox{$\sim$}}\;}

\newcommand{\beq}{\begin{eqnarray}}
\newcommand{\eeq}{\end{eqnarray}}
\def\lsim{\;\raise0.3ex\hbox{$<$\kern-0.75em\raise-1.1ex\hbox{$\sim$}}\;}
\def\gsim{\;\raise0.3ex\hbox{$>$\kern-0.75em\raise-1.1ex\hbox{$\sim$}}\;}

\def\alf{Alfv\'en~}

\def\cmc{\rm ~cm^{-3}}

\def\kms{\rm ~km~s^{-1}}

\def\cmc{\rm ~cm^{-3}}

\def \kms {\rm ~km~s^{-1}}

\def\arcsec{\hbox{$^{\prime\prime}$}}

\def \chan {{\sl Chandra }}

\def \hcm {\hbox {\ifmmode $ atom cm$^{-2}\else atom cm$^{-2}$\fi}}

\def \arcsec {\hbox{$^{\prime\prime}$} }

\def\approxgt{\mathrel{\hbox{\rlap{\lower.55ex \hbox {$\sim$}}
        \kern-.3em \raise.4ex \hbox{$>$}}}}
\def\approxlt{\mathrel{\hbox{\rlap{\lower.55ex \hbox {$\sim$}}
        \kern-.3em \raise.4ex \hbox{$<$}}}}

\def \RXJ1713 {RXJ1713.72--3946 }

\shorttitle{ } \shortauthors{ }

\shorttitle{X-ray Stripes in Tycho's SNR} \shortauthors{Bykov et al}

\begin{document}

\title{X-ray Stripes in Tycho's Supernova Remant: Synchrotron Footprints of
a Nonlinear Cosmic Ray-driven Instability}

\author{Andrei M. Bykov$^1$}
  \affil{Ioffe Physical-Technical Institute, 194021, St. Petersburg, Russia}
  \email{byk@astro.ioffe.ru}

\author{Donald C. Ellison}
  \email{don\_ellison@ncsu.edu}
  \affil{North Carolina State University, Department of Physics, Raleigh, NC 27695-8202, USA}

\author{Sergei M. Osipov$^1$}
  \affil{Ioffe Physical-Technical Institute,  194021, St. Petersburg, Russia}
  \email{osm@astro.ioffe.ru}

  \author{George G. Pavlov$^1$}
  \affil{525 Davey Laboratory, Pennsylvania State
University, University Park,
 PA 16802}
  \email{pavlov@astro.psu.edu}
\altaffiltext{1}{also State Politechnical University, St.\
Petersburg, Russia}

\author{Yury A. Uvarov$^1$}
  \affil{Ioffe Physical-Technical Institute, 194021, St. Petersburg, Russia}
  \email{uv@astro.ioffe.ru}

\date{submitted on the April 28, 2011}


\begin{abstract}
High-resolution \chan observations of Tycho's SNR have revealed
several sets of quasi-steady,  high-emissivity, nearly-parallel
X-ray stripes in some localized regions of the SNR.
These stripes are most likely the result of cosmic-ray (CR) generated
magnetic turbulence at the SNR blast wave. However, for the
amazingly regular pattern of these stripes to appear requires the
simultaneous action of a number of shock-plasma phenomena and is not
predicted by most models of magnetic field amplification. A
consistent explanation of these stripes  yields information on the
complex \NL\ plasma processes connecting efficient CR acceleration
and magnetic field fluctuations in strong collisionless shocks.
The \NL\ diffusive shock acceleration (NL-DSA) model described here,
which includes magnetic field amplification from a  cosmic-ray
current driven instability, does  predict stripes consistent with the
\syn\ observations of Tycho's SNR. We argue that the local ambient
mean magnetic field geometry determines the orientation of the
stripes
and therefore it can be reconstructed with the high resolution X-ray
imaging. The estimated  maximum energy of the CR protons responsible
for the stripes is $\sim 10^{15}$\,eV.
Furthermore, the model predicts that a specific X-ray polarization
pattern, with a polarized fraction $\sim$ 50\%,  accompanies the
stripes, which can be tested with future X-ray polarimeter missions.
\end{abstract}


\section{Introduction}
\label{s_intro}
X-ray synchrotron emission structures have been observed with the
superb spatial resolution of the \chan telescope in  many young
supernova remnants (SNRs) \citep[see
e.g.,][]{vl03,bambaea05,uchiyamaea07,pf08,Eriksen11}.
The morphology of the extended, nonthermal, thin filaments observed
at the SNR edges, and their X-ray brightness profiles, strongly
support the interpretation that $\gsim 10$\, TeV electrons are
accelerated at the forward shock of the expanding supernova shell
and produce \syn\ radiation in an amplified magnetic field.
Recently, very unusual structures consisting of ordered sets of
bright, non-thermal stripes were discovered  by \citet{Eriksen11} in
SN 1572 (Tycho's SNR) with a deep \chan\ exposure. Understanding
these structures presents a formidable challenge for current models
of X-ray \syn\ images of young SNRs.

While diffusive shock acceleration (DSA) has long been favored as
the likely acceleration mechanism for  producing the highly \rel\
electrons needed for X-ray \syn\ emission
\citep[e.g.,][]{be87,je91,md01,EBB2000}, a number of basic questions
remain concerning the nature of DSA in SNRs. One is the relative
efficiency for accelerating protons versus electrons and their
dependence on the magnetic field obliquity, another is the origin of
the strong magnetic field fluctuations required for DSA to produce
CR ions up to the ``knee" near $10^{15}$\,eV in SNRs.
\citep[e.g.,][]{bell78,be87,bk88,BAC2007,vbe08,ber11}.
These questions are all interrelated in non-linear diffusive shock
acceleration (NL-DSA)  and a consistent modeling of the X-ray
stripes seen in Tycho may help resolve  these
 problems.

A fast and efficient mechanism for amplifying magnetic field
fluctuations with scales below the gyroradii of the energetic
particles producing the amplification was proposed by
\citet{bell04,bell05}. Recently, it has been shown that this short-wavelength
turbulence can ignite another instability that produces
turbulence with wavelengths well above the  gyroradii of the
responsible CR particles \citep[][]{boe11}. Large-scale turbulence
is particularly important for determining the maximum energy CRs that a
given shock can produce.
Particle-in-cell (PIC) simulations
\citep[e.g.,][]{niemiecea08,ohiraea09,rs10} have reproduced
the basic predictions of \citet{bell04}.
Studies of the fast Bell instability accounting for the nonlinear
MHD cascade of the growing modes in a parallel shock precursor were
performed by \citet[][]{bell04,zpv08,rsdk08,ze10}.\footnote{In a
strong, CR-modified shock precursor, where the magnetic field
fluctuations likely exceed the mean field value, the distinction
between a parallel and perpendicular shock is blurred. We define a
perpendicular shock as one where the mean field outside the
precursor region is transverse to  the  local shock normal. This
mean ambient  field outside the precursor determines the anisotropy
and polarization of the most strongly amplified modes within the
precursor. Modes with wavevectors along the mean field are most
strongly amplified.}
This work
assumed a CR current as a fixed external
parameter and  showed a fairly
broad distribution of the magnetic fluctuations due to the nonlinear
cascading.
The backreaction of the energetic particle on the shock structure
and the turbulence generation when a sizable fraction of the shock
energy goes  into \rel\ protons is not included in these studies.

To account for the CR current backreaction on the fluctuation
amplification, \citet{vbe09} included  Bell's nonresonant, CR
current driven instability in a \MC\ model of strong shocks
undergoing  efficient particle acceleration.
%
The \MC\ simulation \SCly\ models four strongly  coupled shock
properties: the bulk plasma flow, the full particle spectrum, the
self-generated MHD turbulence including cascading, and particle
injection. In a strong shock, CRs are produced with enough energy
density to modify the upstream flow speed. These precursor CRs
produce a current that results, via the Bell instability, in the
growth of magnetic turbulence. The turbulence, in turn, sets the
momentum and space dependence of the particle diffusion coefficient
and, subsequently, the injection and acceleration efficiency of the
CRs, closing the system.

\citet{vbe09} found  that the shock structure
depends critically on the efficiency of the magnetic turbulence
cascading.\footnote{Only energy transfer from long to short
wavelengths is included in \citet{vbe09}.}
If the cascading along the mean field is suppressed
\citep[e.g.,][]{gs97,bn11} and magnetic field amplification (MFA) is
strong, then the CR shock precursor becomes stratified, and the
turbulence spectrum contains several discrete peaks well separated
in wavenumber, $k$.
We show below that these relatively narrow  peaks, where the wave
spectrum energy density $kW$ can be orders of magnitude greater than
between the peaks \citep[see Figure 3 in][]{vbe09}, can produce
stripes in \syn\ emission consistent with the Tycho SNR
observations. The peaks in turbulence energy density occur because
of the strong coupling between the particle acceleration and MFA
processes. They will not appear  in a test-particle calculation that
ignores this coupling.
In  \citet{vbe09}, three downstream peaks were found for the case
with $\Emax \sim 100$\,TeV. In other examples without cascading, two
and four peaks were found with correspondingly lower and higher
$\Emax$.
We note that while the simulations of \citet{vbe09} were for a
parallel shock, the Bell instability always produces fastest wave
growth along the mean field direction, even though the CR current,
which comes from the CR density gradient, may be in a different
direction \citep[][]{bell05}. The shock modification and
concentration of wave energy into peaks we require should be nearly
independent of the mean field geometry in all strong shocks that
show efficient CR acceleration and large amplitude turbulence as
long as cascading is suppressed. Furthermore, the details of the
peak structure found by \citet{vbe09} are less important for
producing strips than the fact that a narrow, long-wavelength peak
is generated by the maximum energy ions accelerated by the shock.

A satisfactory explanation of the highly regular, nearly-parallel,
X-ray stripes observed in Tycho's SNR by \chan \citep[][]{Eriksen11}
requires a number of conditions to be satisfied. The conditions are:
\newlistroman
\listromanDE the mean magnetic field geometry at the outer blast wave,
where the stripes are prominent, must be nearly perpendicular, i.e.,
the field is perpendicular to the local shock normal;
\listromanDE the unstable growing MHD modes must be linearly
polarized and maintain coherence over a fairly long spatial
scale, $l_{\rm c}$, that is somewhat below the scale size of the CR
precursor,
%
\listromanDE the background turbulence must have narrow peaks in
wavenumber $k$;
\listromanDE  the stripes must persist without significant variation
long enough to be seen in a deep \chan observation;
\listromanDE the \rel\ electrons producing the \syn\ radiation must
have a steep spectrum  to enhance the emissivity contrast between
regions of low and high magnetic field;
\listromanDE NL-DSA must be efficient enough in a
quasi-perpendicular configuration to produce a shock precursor
structure and MFA; and
\listromanDE turbulence cascading along the mean field should be
suppressed to prevent spectral broadening.
The model we now detail  provides all  of these properties in rather a
natural way.

\section{Model}
The coherent nature of the X-ray stripes  suggests that
the underlying magnetic turbulence is strongly anisotropic. Isotropic turbulence
would not produce extended coherent structures
with thin stripes.
Both the Bell short-wavelength instability and the long-wavelength
instability, considered by \citet{boe11},  produce anisotropic
turbulence with a prominent growth-rate maximum along the mean
ambient magnetic field direction.
The mean ambient magnetic field is the field
averaged over scales larger than the CR shock precursor.
Since the
SNR radius in Tycho is considerably larger than the shock precursor, the local
ambient field direction may vary over the SNR surface.

\begin{figure*}[t]
\includegraphics[width=0.9\textwidth,angle=0]{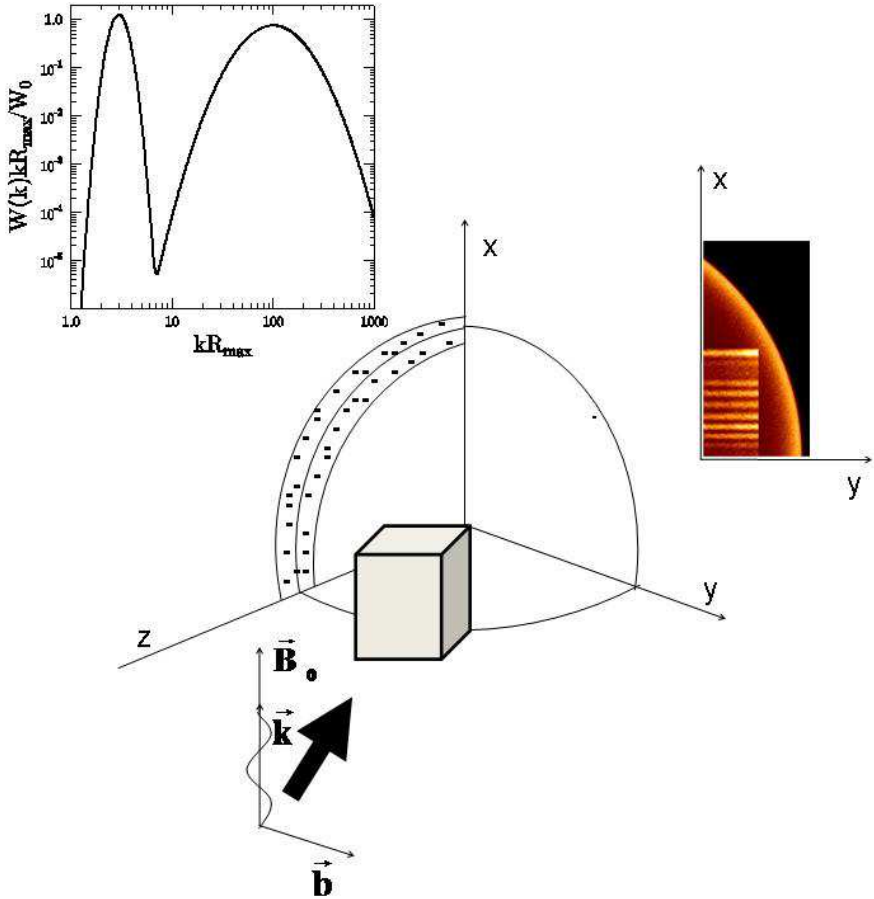}
\caption{Geometry of the simulated supernova shell. The central
figure  shows half of the shell quarter. The upstream magnetic field
amplification region, where the CR-driven instability occurs, is
shown as a box with a scale size that is about the coherence length
of the amplified field $l_{\rm c}$. The local mean magnetic field in
the upstream flow is perpendicular to the flow velocity indicated as
the thick arrow. The wavevector of the amplified mode is along the
mean field, and the linearly polarized MHD-wave (in the
$y$-direction) has amplitude $|\vec{b}|$. The simulated spectrum of
the magnetic fluctuations is shown in the left inset. We show in the
inset the turbulence spectrum, $W(k)k\rmg/W_0$ where the
normalization factor $W_0$ is expressed through the r.m.s.
downstream magnetic field as $W_0 = \rmg \left < B^2 \right
>/(16\pi)$. The dotted area in the central figure shows a
cross-section of the spherical shell filled with the accelerated
X-ray radiating electrons. The local densities of the Stokes
parameters were integrated over the line of sight along the $z$-axis
producing the image illustrated in the right inset and in
Fig.~\ref{image}.} \label{geom}
\end{figure*}

\begin{figure*}[t]
\includegraphics[width=0.60\textwidth]{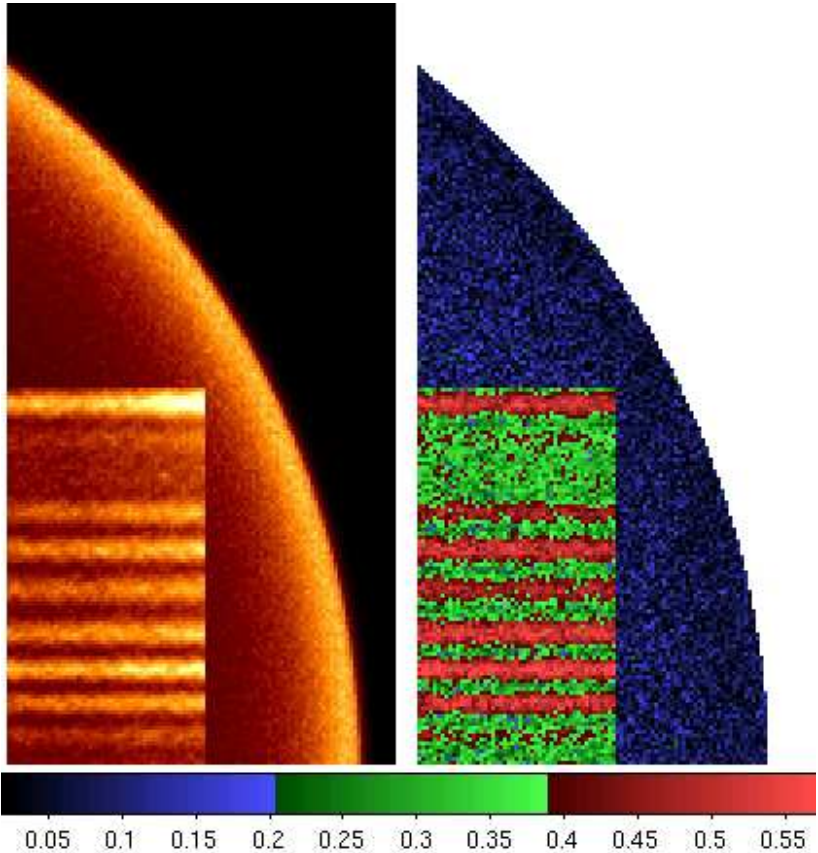}
\caption{Supernova remnant synchrotron emission images simulated in
the NL-DSA model, accounting for magnetic field amplification from a
CR current driven instability. The left panel is the synchrotron
X-ray intensity at 5 keV. The degree of polarization of the X-ray
emission is shown in the right panel, and the values correspond to
the color bar. The relatively high polarization fraction is mainly
due to the peaked structure of the magnetic fluctuation spectrum and
the steepness of the distribution of \syn\ emitting electrons.}
\label{image}
\end{figure*}

The SNR geometry we model is shown in Figure~\ref{geom}, where the
ambient (i.e., upstream) field, $\vec{B_0}$, is in the $x$-direction
and is tangent to the shock  surface for the region of the forward
shock in the $yz$-plane. The pattern shown has the linear
polarization vector, $\vec{b}$,  in the $y$-direction and the
wavevectors of the growing modes are along $\vec{B_0}$.
 The rectangular box schematically shows
the magnetic field amplification region where the CR-driven
instability occurs. The inflowing fluctuations in the ISM field with
wavevectors along the mean field direction are amplified most
rapidly by the CR-current instability in the box of scale size
$l_{\rm c}$. This scale determines the coherence length of the
amplified field. The projected scale of the box is imprinted into
the simulated image shown in the top-left inset in Figure~\ref{geom}
and in Figure~\ref{image}. The linear polarization $\vec{b}$ of the
waves propagating along the downstream field of the transverse shock
results in regions parallel to the shock surface with alternating
high and low field strengths,  which will translate into bright and
dim regions of \syn\ emission.

In both the Bell and long-wavelength
instabilities, the polarization of the turbulence depends on
the orientation of the ambient magnetic field
\citep[see][]{bell05,boe11}.
The amplified field
fluctuations have circular polarization for a parallel shock
where the ambient field and the CR density gradient are
directed along the shock normal.
In contrast, linearly polarized waves, propagating along the ambient
field, are amplified in the case of a perpendicular shock.
The mode polarization is important to produce stripes as well as the
shock geometry. If the stripes were produced in the quasi-parallel
shock, they would be limited in length to the thickness of the \syn\
-bright downstream region, $\delR$, and this thickness is estimated
below to be small compared to the length of the observed stripes.
The linearly polarized waves in a perpendicular shock, however,  can
produce a nearly static, striped emission pattern with a length much
longer than $\delR$  along the shock surface.

However, even with linearly polarized waves in a uniform
perpendicular field, any stripes that might occur will be washed out
if the waves in the long-wavelength peak have a significant spread
in wave number, i.e., if $\delK > k_0$, where $k_0$ is the
long-wavelength peak position in the two peak spectrum shown in the
inset in Figure~\ref{geom}.\footnote{Fluctuations with wavelengths
longer than produced by the Bell instability may also be amplified
\citep[e.g.,][]{boe11} or produced by inverse cascading
\citep[]{rs10}.  The longer wavelengths have slower growth rates
then those produced by the fast Bell instability.  These may
somewhat broaden the spectral peak. Any longer wavelength turbulence
will increase the maximum CR proton energy, and if this turbulence
is included in the Monte Carlo simulation, there would be a shift of
$k_0$ to smaller $k$. A NL-DSA model that  accounts for all of the
resonant and nonresonant instabilities is not yet available.}
The peaks found by \citet{vbe09} obey $\delK < k_0$. When combined
with linearly polarized waves in a perpendicular shock, these narrow
peaks  result in the creation of stripe-like structures in the
synchrotron image.

We constructed synchrotron images of a spherical SNR in a model that
accounts for the fluctuating magnetic field structures described
above following the approach developed in \citet{bue08,bubhk09}. The
amplified magnetic field in the downstream is simulated as a
superposition of the predominantly linearly polarized modes of random
phases, with a spectral distribution consisting of a few isolated
narrow peaks in wavenumber space, as was obtained in the
NL-DSA model with a CR-driven instability.
The magnetic fluctuation spectrum, shown in the inset in
Fig.~\ref{geom}, has two peaks; the long-wavelength peak at $k_{\rm
0} \sim 2\pi\rmg^{-1}$  (where $\rmg$ is the gyro-radius of the
maximum energy ion  accelerated by the shock) has a width $\delK
\approx 0.5k_{\rm 0}$. The second peak at the shorter wavelength
(with $k_1 \sim 33\times k_0$) is wider. The energy density of
magnetic fluctuations in both peaks are comparable, as was the case
in the NL-DSA model by \citet{vbe09}. The scales corresponding to
the short-wavelength peak at $k_1$ cannot be resolved in Tycho's SNR
even with arcsecond resolution X-ray telescopes, while the
long-wavelength fluctuations have scales resolvable with these
instruments.

Regarding the timing and spacing of the stripes, the wavelength of
the turbulence at $\kpeak$ must be consistent with the observed
spacing, and the period must be long enough so that the structure
does not change significantly in a deep \chan observation of
duration $\dTobs$.

In the quasi-perpendicular portion of the SNR
shock the stripes move transverse to the flow velocity.
The structure moves  with a phase velocity on the order of the
\alf velocity $v_a$ and the time scale for changes in the structure
is $\sim (\kpeak v_a)^{-1}$, yielding a constraint $(\kpeak
v_a)^{-1}
> \dTobs$. The separation of the stripes is determined by the
wavelength $\sim 1/\kpeak$. We estimate that the apparent movement
of the stripes due to this effect is $<0.02\arcsec$/yr for a SNR at
$\sim 4$\,kpc, expanding in ambient gas of number density $\sim 0.1
\cmc$, and assuming the amplified magnetic field downstream from the
shock is $\sim 60$\,\muG, yielding $v_a \sim 450 \kms$. Given the
extended spatial structure of the stripes, their proper motion can
hardly be noticed in 10 years of \chan observations.

An important point to address is why the stripes that
appear in the X-ray \syn\ emission do not appear in the radio \syn\
observations of Tycho's SNR
\citep[][]{ReynosoEtal1997,Vighetal2011}. This is naturally
explained if the $>10$\,TeV electrons producing the X-rays have a
steeper  distribution than the $\sim$\,GeV electrons producing the
radio. The
 synchrotron emissivity of a power-law electron spectrum with
 spectral index $\alpha$ is proportional to $B^{(\alpha + 1)/2}$, i.e.,
 the local emissivity is relatively very high for large local $B$ and large
$\alpha$. With a steep electron distribution, the intensity contrast
between the peak and trough of the magnetic turbulence produces a
greater luminosity contrast for the $>$\,TeV electrons than for the
flatter $\sim$\,GeV electrons. The stripes in radio are not distinct
enough to be visible in the VLA observations.

The kinetic model we used here to simulate the electron distribution
is similar to that discussed by \citet{bubhk09}. The spatially
inhomogeneous electron distribution function is calculated from the
kinetic equation for electrons at a SNR shock using a piece-wise
parametrization of the particle diffusion coefficient  that is
consistent with that obtained for $\sim 10$\,TeV protons by
\citet{vbe09}. The synchrotron-Compton losses of electrons in
magnetic fields of $\sqrt{\left < B^2 \right >}$  are
accounted for. The narrow spherical shell of thickness $\delR$ filled
with the accelerated electrons is shown schematically in
Figure~\ref{geom} as a dotted region.

Finally, as \citet{vbe09} showed, energy cascading  spreads out the
peaks in turbulence power and eliminate the key feature needed to
produce an ordered pattern of stripes. Therefore, such stripes can
form only if the turbulent cascading along the mean magnetic field
is quenched.

\section{Discussion}
The width $\delR$  of the region with \syn\ emitting electrons in
the shock downstream is typically rather narrow (dotted region in
Figure~\ref{geom}).
The thickness of the layer depends on the cooling rate in  the
downstream magnetic field:
$\delR \sim 6
\times 10^{15} \cdot v_{\rm sh8}\cdot B_{\rm mG}^{-2}\cdot
\ETeV^{-1}$ cm,
where the r.m.s. magnetic field $B_{\rm mG}$ is in $m$G,  $v_{\rm
sh8}$ is the shock velocity in units of $10^3\,\kms$, and $\ETeV$ is
the electron energy in TeV.
After accounting for the projection effect, the value of $\delR \ll
R$ provides the limb brightening observed in many synchrotron images
of young SNRs.

While the synchrotron X-ray emission is mainly produced in a thin
downstream layer that cannot always be resolved in X-rays, the
transverse (relative to the shock normal)  scale size of the
emitting region is determined by the projected scale size of the
magnetic field amplification region in the shock precursor. The wave
vectors of the fast growing modes flowing into the shock  must be
directed, in the shock rest frame, within a fairly narrow cone along
the mean ambient field direction.
The transverse scale size of the locally amplified fluctuations
should be comparable to the precursor scale size, which is assumed
to be much smaller than the shock radius.
It is also important that the fast amplification of the field
implies that the field correlation length $l_{\rm c}$ is also on
the order of
the precursor scale size. Therefore, the length of a stripe is
determined by the projected scale size of the magnetic field
amplification region.
In the model under consideration, the fast amplification of magnetic
field fluctuations due to the CR current-driven instability occurs
in the CR precursor of the strong shock. The scale size of the
CR-precursor, for Bohm diffusion,  is a fraction $f$ of $c/v_{\rm
sh} \times \rmg$, and  typically $f\sim 0.1-0.3$
from the Monte Carlo simulations.

The distance between the stripes is set by the wavelength  of the
turbulence which is on the order of $\rmg$. Therefore, the ratio of
the stripe elongation length to the distance between the field maxima
is about  $f c/v_{\rm sh}$, and that number also estimates the
number of the stripes in a single coherent set shown in
Figure~\ref{image}.

\section{Conclusions}
It has been known for quite some time that NL-DSA can be efficient with a strong coupling
between the CR spectrum and the magnetic field dynamics.
However, a \NL\ study, using \MC\ techniques by \citet[][]{vbe09},
has recently  produced a surprising result: if the turbulence
cascading is suppressed along the mean magnetic field, the shock
precursor becomes stratified, and a multiple-peak structure of the
amplified, fluctuating magnetic field spectrum forms.

As we have shown here, the multiple-peak power spectrum, along with
the anisotropic character of the CR-current driven instability, can
result in structures in synchrotron images of SNRs similar to
the recent observations of Tycho's SNR by \citet[][]{Eriksen11}.
The unusual nature of the nearly-parallel X-ray stripes is highly
constraining and a number of conditions must be fulfilled for them
to be produced.
Observationally, the stripe-like structures should be prominent and
resolvable for a quasi-perpendicular shock configuration, but they
are likely  too small or indistinct to be resolved with current
instruments in regions where the shock is quasi-parallel. The
synchrotron emission from the stripes should be linearly polarized.
The simulated degree of polarization is shown in the right panel in
Figure~\ref{image}.
The high polarized fraction of $\sim 50\%$ predicted for the bright
stripes should make these structures highly prominent in future X-ray
polarization observations.  The lack of Faraday rotation and the
relatively steep spectra of synchrotron emission are advantageous
for  X-ray polarization studies of young SNRs \citep[see][]{bubhk09}.

Stripe-like structures should form in a section where the local
field lies along the shock surface and where the turbulence
cascading is suppressed.
We argue that it is difficult to produce the stripe structure in
synchrotron images in any natural way other than with narrow peaks
in the magnetic turbulence in a perpendicular shock. This uniqueness
offers a new way  to infer the geometry of the local ambient
magnetic field. The magnetic field structure is particularly
important in modeling the TeV gamma-ray emission from Tycho's SNR
recently detected with the {\sl VERITAS} ground based gamma-ray
observatory by \citet{veritas11}.

\begin{acknowledgements}
  We thank the anonymous referee for constructive comment.
 A.M.B., S.M.O., G.G.P and Y.A.U were supported in part by
 the Russian government grant 11.G34.31.0001 to Sankt-Petersburg
 State Politechnical University,
and by the RAS Presidium Program and RBRF OFIm 11-02-12082. The
simulations were performed at the Joint Supercomputing Center JSCC
RAS and the Supercomputing Center at Ioffe. A.M.B and D.C.E. were
supported by NSF PHY05-51164. D.C.E. was supported by NASA grants
ATP02-0042-0006, NNH04Zss001N-LTSA, and 06-ATP06-21. G.G.P was
supported in part by NASA grant NNX09AC84G.

\end{acknowledgements}


\begin{thebibliography}{30}
\expandafter\ifx\csname
natexlab\endcsname\relax\def\natexlab#1{#1}\fi

\bibitem[{{Acciari}{et~al.}(2011)}]{veritas11}{Acciari}, V.~A.~et~al. 2011, \apjl,
  730, L20+

\bibitem[{{Bamba} {et~al.}(2005){Bamba}, {Yamazaki}, {Yoshida}, {Terasawa}, \&
  {Koyama}}]{bambaea05}
{Bamba}, A., {Yamazaki}, R., {Yoshida}, T., {Terasawa}, T., \&
{Koyama}, K.
  2005, \apj, 621, 793

\bibitem[{{Bell}(1978)}]{bell78}
{Bell}, A.~R. 1978, \mnras, 182, 147

\bibitem[{{Bell}(2004)}]{bell04}
---. 2004, MNRAS, 353, 550

\bibitem[{{Bell}(2005)}]{bell05}
---. 2005, \mnras, 358, 181

\bibitem[{{Berezhko} \& {Krymski{\u i}}(1988)}]{bk88}
{Berezhko}, E.~G., \& {Krymski{\u i}}, G.~F. 1988, Soviet Physics
Uspekhi, 31,
  27
\bibitem[{{Blandford} \& {Eichler}(1987)}]{be87}
{Blandford}, R., \& {Eichler}, D. 1987, Phys. Rep., 154, 1

\bibitem[{{Blasi} {et~al.}(2007){Blasi}, {Amato}, \& {Caprioli}}]{BAC2007}
{Blasi}, P., {Amato}, E., \& {Caprioli}, D. 2007, \mnras, 375, 1471

\bibitem[{{Brandenburg} \& {Nordlund}(2011)}]{bn11}
{Brandenburg}, A., \& {Nordlund}, {\AA}. 2011, Reports on Progress
in Physics,
  74, 046901

\bibitem[{{Bykov} {et~al.}(2011{\natexlab{a}}){Bykov}, {Ellison}, \&
  {Renaud}}]{ber11}
{Bykov}, A.~M., {Ellison}, D.~C., \& {Renaud}, M.
2011{\natexlab{a}}, \ssr, DOI: 10.1007/s11214-011-9761-4

\bibitem[{{Bykov} {et~al.}(2011{\natexlab{b}}){Bykov}, {Osipov}, \&
  {Ellison}}]{boe11}
{Bykov}, A.~M., {Osipov}, S.~M., \& {Ellison}, D.~C.
2011{\natexlab{b}},
  \mnras, 410, 39

\bibitem[{{Bykov} {et~al.}(2009){Bykov}, {Uvarov}, {Bloemen}, {den Herder}, \&
  {Kaastra}}]{bubhk09}
{Bykov}, A.~M., {Uvarov}, Y.~A., {Bloemen}, J.~B.~G.~M., {den
Herder}, J.~W.,
  \& {Kaastra}, J.~S. 2009, \mnras, 399, 1119

\bibitem[{{Bykov} {et~al.}(2008){Bykov}, {Uvarov}, \& {Ellison}}]{bue08}
{Bykov}, A.~M., {Uvarov}, Y.~A., \& {Ellison}, D.~C. 2008, \apjl,
689, L133

\bibitem[{Ellison {et~al.}(2000)Ellison, Berezhko, \& Baring}]{EBB2000}
Ellison, D.~C., Berezhko, E.~G., \& Baring, M.~G. 2000, ApJ, 540,
292

\bibitem[{{Eriksen} {et~al.}(2011){Eriksen}, {Hughes}, {Badenes}, {Fesen},
  {Ghavamian}, {Moffett}, {Plucinksy}, {Rakowski}, {Reynoso}, \&
  {Slane}}]{Eriksen11}
{Eriksen}, K.~A., {Hughes}, J.~P., {Badenes}, C., {Fesen}, R.,
{Ghavamian}, P.,
  {Moffett}, D., {Plucinksy}, P.~P., {Rakowski}, C.~E., {Reynoso}, E.~M., \&
  {Slane}, P. 2011, Astrophysical Journal, 728, L28+

\bibitem[{{Goldreich} \& {Sridhar}(1997)}]{gs97}
{Goldreich}, P., \& {Sridhar}, S. 1997, \apj, 485, 680

\bibitem[{{Jones} \& {Ellison}(1991)}]{je91}
{Jones}, F.~C., \& {Ellison}, D.~C. 1991, Space Science Reviews, 58,
259

\bibitem[{{Malkov} \& { Drury}(2001)}]{md01}
{Malkov}, M.~A., \& { Drury}, L. 2001, Reports on Progress in
Physics, 64, 429

\bibitem[{{Niemiec} {et~al.}(2008){Niemiec}, {Pohl}, {Stroman}, \&
  {Nishikawa}}]{niemiecea08}
{Niemiec}, J., {Pohl}, M., {Stroman}, T., \& {Nishikawa}, K. 2008,
\apj, 684,
  1174

\bibitem[{{Ohira} {et~al.}(2009){Ohira}, {Reville}, {Kirk}, \&
  {Takahara}}]{ohiraea09}
{Ohira}, Y., {Reville}, B., {Kirk}, J.~G., \& {Takahara}, F. 2009,
\apj, 698,
  445

\bibitem[{{Patnaude} \& {Fesen}(2009)}]{pf08}
{Patnaude}, D.~J., \& {Fesen}, R.~A. 2009, \apj, 697, 535

\bibitem[{{Reville} {et~al.}(2008){Reville}, {O'Sullivan}, {Duffy}, \&
  {Kirk}}]{rsdk08}
{Reville}, B., {O'Sullivan}, S., {Duffy}, P., \& {Kirk}, J.~G. 2008,
\mnras,
  386, 509

\bibitem[{{Reynoso} {et~al.}(1997){Reynoso}, {Moffett}, {Goss}, {Dubner},
  {Dickel}, {Reynolds}, \& {Giacani}}]{ReynosoEtal1997}
{Reynoso}, E.~M., {Moffett}, D.~A., {Goss}, W.~M., {Dubner}, G.~M.,
{Dickel},
  J.~R., {Reynolds}, S.~P., \& {Giacani}, E.~B. 1997, \apj, 491, 816

\bibitem[{{Riquelme} \& {Spitkovsky}(2010)}]{rs10}
{Riquelme}, M.~A., \& {Spitkovsky}, A. 2010, \apj, 717, 1054

\bibitem[{{Uchiyama} {et~al.}(2007){Uchiyama}, {Aharonian}, {Tanaka}, \&
  et~al.}]{uchiyamaea07}
{Uchiyama}, Y., {Aharonian}, F.~A., {Tanaka}, T., \& et~al. 2007,
\nat, 449,
  576

\bibitem[{{Vigh} {et~al.}(2011){Vigh}, {Vel{\'a}zquez}, {G{\'o}mez}, {Reynoso},
  {Esquivel}, \& {Matias Schneiter}}]{Vighetal2011}
{Vigh}, C.~D., {Vel{\'a}zquez}, P.~F., {G{\'o}mez}, D.~O.,
{Reynoso}, E.~M.,
  {Esquivel}, A., \& {Matias Schneiter}, E. 2011, \apj, 727, 32

\bibitem[{{Vink} \& {Laming}(2003)}]{vl03}
{Vink}, J., \& {Laming}, J.~M. 2003, \apj, 584, 758

\bibitem[{{Vladimirov} {et~al.}(2008){Vladimirov}, {Bykov}, \&
  {Ellison}}]{vbe08}
{Vladimirov}, A.~E., {Bykov}, A.~M., \& {Ellison}, D.~C. 2008, \apj,
688, 1084

\bibitem[{{Vladimirov} {et~al.}(2009){Vladimirov}, {Bykov}, \&
  {Ellison}}]{vbe09}
---. 2009, \apjl, 703, L29

\bibitem[{{Zirakashvili} {et~al.}(2008){Zirakashvili}, {Ptuskin}, \&
  {V{\"o}lk}}]{zpv08}
{Zirakashvili}, V.~N., {Ptuskin}, V.~S., \& {V{\"o}lk}, H.~J. 2008,
\apj, 678,
  255

\bibitem[{{Zweibel} \& {Everett}(2010)}]{ze10}
{Zweibel}, E.~G., \& {Everett}, J.~E. 2010, \apj, 709, 1412

\end{thebibliography}

\end{document}